\documentclass[a4paper,10pt,twocolumn]{elsarticle}

\usepackage{mathrsfs,amsmath,amssymb,multicol}
\usepackage[numbers]{natbib}
\usepackage{color}

\biboptions{numbers,sort&compress}
\journal{Computers $\&$ fluids}

\begin{document}

\title{ Efficient lattice Boltzmann models for the Kuramoto-Sivashinsky equation}


\author{Hiroshi Otomo}\corref{mycorrespondingauthor}\ead{hiroshi.otomo@tufts.edu}\author{Bruce M. Boghosian}
\address{Department of Mathematics, Tufts University, Medford, Massachusetts 02155, USA}

\author{Fran\c{c}ois Dubois}
\address{CNAM Paris, Laboratoire de m\'{e}canique des structures et des syst\`{e}mes coupl\'{e}s,\\ 292, rue Saint-Martin, 75141 Paris cedex 03,France}
\address{Universit\'{e} Paris-Sud, Laboratoire de math\'{e}matiques, UMR CNRS 8628, 91405 Orsay cedex, France}
\address{Department of Mathematics, University Paris-Sud, Bat. 425, F-91405  Orsay, France}


\twocolumn[
  \begin{@twocolumnfalse}

\begin{abstract}
In this work, we improve the accuracy and stability of the lattice Boltzmann model for the Kuramoto-Sivashinsky equation proposed in \cite{2017_Otomo}.  This improvement is achieved by controlling the relaxation time, modifying the equilibrium state, and employing more and higher lattice speeds, in a manner suggested by our analysis of the Taylor-series expansion method.  The model's enhanced stability enables us to use larger time increments, thereby more than compensating for the extra computation required by the high lattice speeds.  Furthermore, even though the time increments are larger than those of the previous scheme, the same level of accuracy is maintained because of the smaller truncation error of the new scheme.  As a result, total performance with the new scheme on the D1Q7 lattice is improved by 92 $\%$ compared to the original scheme on the D1Q5 lattice.
\end{abstract}
\maketitle
\end{@twocolumnfalse}

]

%
%

\section{Introduction}

The Kuramoto-Sivashinsky (KS) equation is well known to reproduce a variety of chaotic phenomena caused by intrinsic instability such as the unstable behavior of laminar flame fronts \cite{1980_Sivashinsky, 2015_Sharafatmandjoor}, thin-water-film flow on a vertical wall \cite{1977_Sivashinsky}, and persistent wave propagation through a reaction-diffusion system \cite{1976_Kuramoto}. 
For space $X$ and time $T$, the KS equation for a quantity $\rho$ is
\begin{equation}
\label{KS_eq}
\partial_T \rho + \rho \partial_{X} \rho = -\partial^2_{X} \rho-\partial^4_{X} \rho.
\end{equation}
The second term on the left-hand side is the nonlinear advection term, while the first and second  terms on the right-hand side are the production and hyperdiffusion terms, respectively.  Examining the relationship between those terms, Holmes~\cite{2012_Holmes} found that the KS equation exhibits basic properties of turbulent flow, and indeed corresponds to the equation for the fluctuating velocity derived from the Navier-Stokes equation.  Accordingly, the KS equation is often used to explore basic features of chaotic systems.

The lattice Boltzmann (LB) method was originally developed from models of lattice-gas cellular automata, and is based on principles of kinetic theory~\cite{2001_Succi}.  The ensemble of particle states is described by a distribution function which evolves through the particles' advection and collision process, thereby establishing the hydrodynamics.  In addition to the hydrodynamic degrees of freedom, the model's kinetic modes depend on higher moments of the distribution function and give rise to peculiar features of the LB method, which are also beneficial for more detailed numerical modeling.  

In the last decade, a number of LB models for nonlinear spatiotemporal systems have been developed~\cite{2009_Huilin,2009_Baochang,2009_Zhang,2010_Jie,2009_Yan,2011_Lina,2017_Otomo}.  In a previous study~\cite{2017_Otomo}, LB models for nonlinear equations, such as the Burgers', Korteweg-de Vries, and Kuramoto-Sivashinsky (KS) equations, were derived using both the Chapman-Enskog and Taylor-series expansion methods~\cite{2004_Holdych,2008_Dubois} consistently.  For simulating the long-time behavior of these chaotic equations accurately, however, the LB models thus derived require substantial computational time.  Moreover, whereas the relaxation time $\tau$ in the LB model for the Navier-Stokes equation has a clear relationship to the viscosity, the role and optimized value of $\tau$ for the KS equation is not at all clear, and its value had to be set by trial and error.  In this work, remedies for both of these issues are investigated using the Taylor-series expansion method, which allows for easy analysis of higher-order effects in the hydrodynamic equations.

This paper is organized as follows:  In Section~\ref{sec:LB_formalism}, we present a way to improve the LB model for the KS equation.  In Section~\ref{sec:result}, we test the LB model thereby derived by comparisons with analytic solutions and with the previous model. In Section.~\ref{sec:summary}, we summarize the results of this study and present conclusions.

\section{Improved lattice Boltzmann models for the Kuramoto-Sivashinsky equation}
\label{sec:LB_formalism}

With discrete lattice velocities $c_i$ and the relaxation time $\tau$, the LB equation for the discrete distribution function $f_i$ is given by: 
\begin{equation}
\label{LB_equation_ordinary}
f_{i} \left( x + c_i \Delta t, t+\Delta t \right) -f_{i} \left( x, t \right)= -\frac{f_{i} - f^{eq}_{i}}{\tau}.
\end{equation}
Here $f^{eq}_{i}$ is the local equilibrium state whose form for the KS equation, Eq.~(\ref{KS_eq}), was found in prior work~\cite{2017_Otomo} to be
\begin{equation}
\label{feq}
f^{eq}_{i}=\rho \left( w^{(0)}_{i} +  \mathcal{K} w^{(2)}_{i} +   \mathcal{M} w^{(4)}_{i}  \right) + \rho^2 \mathcal{J} w^{(1)}_{i},
\end{equation}
where $\rho = \sum_i f_i$ and where the weights $w_{i}$ have moments shown in Table.~\ref{tab:moment_our_scheme}.  Explicit forms for these weights are presented in \ref{appendix:weight}.  The quantities $\mathcal{K}$, $\mathcal{M}$, and  $ \mathcal{J}$  are given in Table.~\ref{tab:moment_nonlineareq}, where we have defined $\mathcal{T}_i =\sum_{n=1}^{\infty} \left( 1 - \frac{1}{\tau} \right)^n \left[ \left( n+1 \right)^i -n^i \right]$.  For $\tau >1/2$ these are
\begin{eqnarray}
\label{T_form}
\mathcal{T}_1 &=&  \tau -1 \nonumber\\
\mathcal{T}_2 &=& 2\tau^2 - \tau -1 \nonumber \\
\mathcal{T}_3 &=& 6\tau^3 -6 \tau^2 + \tau -1 \nonumber \\
\mathcal{T}_4 &=& \left( \tau-1 \right) \left( 24 \tau^3-12 \tau^2 +2 \tau +1 \right).
\end{eqnarray}
The characteristic lattice speed $|c|$, which is dimensioned in lattice units, is assumed to be one and not explicitly written in what follows.

\begin{table*}[htbp]
  \begin{center}
    \begin{tabular}{c}
      \begin{minipage}{0.38\hsize}
        \begin{center}
          \tabcolsep = 0.1cm
        \begin{tabular}{c c c c c c } \hline
	Order of moments &  $w^{(0)}_{i}$  &  $w^{(1)}_{i}$     &  $w^{(2)}_{i}$    &     $w^{(4)}_{i}$  \\ \hline
          0      &     1 & 0      &    0          &  0     \\ 
          1      &     0 & 1      &    0          &  0   \\ 
          2      &     0 & 0      &    1          & 0     \\ 
          3      &     0 & 0      &    0         & 0     \\ 
          4      &     0 & 0    &    0           &  1   \\ \hline
	\end{tabular}
        \end{center}
        \caption{Moments of $w_i$ }
        \label{tab:moment_our_scheme}
      \end{minipage}
      \begin{minipage}{0.62\hsize}
        \begin{center}
          \tabcolsep = 0.15cm
        \begin{tabular}{c  c} \hline	   
          $\mathcal{J}$:      &       $\beta /2 \alpha$            \\ 
          $\mathcal{K}$:      &       $  -2 \beta \left(  \mathcal{T}_1 +1  \right)/ \left\{ \alpha^2 \left( \mathcal{T}_2 +1 \right) \right\} $                  \\ 
          $\mathcal{M}$:      &     $- 24 \beta  \left(  \mathcal{T}_1 +1  \right) / \left\{ \alpha^4  \left(  \mathcal{T}_4 +1  \right)  \right\}$    \\ \hline
	\end{tabular}
        \end{center}
        \caption{Coefficients of moments}
        \label{tab:moment_nonlineareq}
      \end{minipage}
    \end{tabular}
  \end{center}
\end{table*}

We use the Taylor-series expansion method for a small non-dimensional parameter $\epsilon$, with the scaling assumptions $\Delta x /L = \epsilon$ and $\Delta t / T = \epsilon^{m}$ for $m>1$, and we assume that $L \partial_x$ and $T \partial_t$ are order unity, where $L$ and $T$ are macroscopic length and time scales.  By summing over $i$ in Eq.~(\ref{LB_equation_ordinary}), one obtains~\cite{2017_Otomo},
\begin{eqnarray}
\label{taylor_result}
\lefteqn{
\frac{\partial \rho}{\partial t} 
=
-\mathcal{J} \frac{\partial \rho ^2 }{\partial x}
+\frac{ \Delta t }{2!} \mathcal{K} \frac{\partial^2 \rho}{\partial x^2} \frac{ \mathcal{T}_2 +1  }{\mathcal{T}_1+1}} \nonumber \\
& &
+\frac{\left( \Delta t \right)^3 }{4!} \mathcal{M} \frac{\partial^4 \rho}{\partial x^4} \frac{ \mathcal{T}_4 +1  }{\mathcal{T}_1+1} 
+ \mathcal{O} \left( \frac{\partial^5 \rho }{\partial x^5}, \frac{\partial^2 \rho^2 }{\partial x \partial t}, \frac{\partial^2 \rho }{ \partial t^2} \right).\nonumber\\
\end{eqnarray}
The last term on the right-hand side is regarded as the truncation error for the KS equation.  

In a simulation, the physical space $X$ and physical time $T$ are scaled with parameters $\alpha$ and $\beta$ from the corresponding coordinates in lattice units, $x$ and $t$, as follows
\begin{eqnarray}
\label{scaling}
X=\alpha x \nonumber \\
T=  \beta t. 
\end{eqnarray}
The increments of physical space and time are therefore $\Delta X = \alpha$ and $\Delta T = \beta$.  Taking this into account, it is straightforward to see that Eq.~(\ref{KS_eq}) can be derived from Eq.~(\ref{taylor_result}) with the choices of $\mathcal{J}$, $\mathcal{K}$ and $\mathcal{M}$ given in Table~\ref{tab:moment_nonlineareq}.
 
\subsection{Strategy}

It is worth highlighting several features of the above formalism:
\begin{itemize}
\item[$\bullet$] The requirements for the weights set forth in Table~\ref{tab:moment_our_scheme} can be satisfied with at least 5 lattice speeds, so D1Q5 would work.
\item[$\bullet$] Higher moments than those shown in Table~\ref{tab:moment_our_scheme} impact only the truncation error in Eq.~(\ref{taylor_result}).
\item[$\bullet$] As long as $\mathcal{K}$, $\mathcal{M}$, and  $ \mathcal{J}$ are as given in Table.~\ref{tab:moment_nonlineareq}, the relaxation time does not influence the leading order terms in  Eq.~(\ref{taylor_result}), but only the truncation error.
\end{itemize}

Due to the first point above, we adopt the D1Q5 lattice in this paper as our ``basic scheme.''
Due to the second and third points, we see that by using more lattice speeds than the basic scheme and by varying the relaxation time, we may enhance accuracy while retaining stability. Although the increased number of speeds will require additional computation, if the time increments for achieving the same accuracy can be increased significantly, the total computational cost will be improved.

\subsection{Analysis}
\label{sec:KS_formalism}

According to our basic scheme, the leading truncation error term at order $\beta^0$ of Eq.~(\ref{taylor_result}) is the sixth spatial derivative term whose coefficients involve $\mathcal{K}$ and $\mathcal{M}$.  By straightforward algebra, we find that this error term is
\begin{eqnarray}
\label{KS_trunc11}
\left\{ \frac{\alpha^4 \left( \mathcal{T}_6+1 \right) }{90 \left( \mathcal{T}_2+1 \right) } 
- \frac{\alpha^2 \left( \mathcal{T}_6+1 \right)  }{6 \left( \mathcal{T}_4+1 \right) } 
\right\} \frac{\partial^6 \rho}{\partial X^6},
\end{eqnarray}
where we have substituted the forms of $\mathcal{K}$ and $\mathcal{M}$ from Table~\ref{tab:moment_nonlineareq}.  Similarly, the truncation error terms at order $\beta$ of Eq.~(\ref{taylor_result}) are those involving $\partial^2 \rho / \partial t^2$, $\partial^3 \rho / \partial t \partial x^2$, and $\partial^5 \rho / \partial t \partial x^4$, whose coefficients also involve $\mathcal{K}$ and $\mathcal{M}$. By utilizing the leading order result, Eq.~(\ref{KS_eq}), these explicit forms are derived as
\begin{eqnarray}
\label{KS_trunc2}
\beta \left\{ \frac{  \mathcal{T}_2+1  }{2 \left( \mathcal{T}_1+1 \right) } 
- \frac{  \mathcal{T}_3+1  }{  \mathcal{T}_2+1  }
\right\} \frac{\partial^4 \rho}{\partial X^4}  \nonumber \\
+ \beta \left\{ \frac{  \mathcal{T}_2+1 }{ \mathcal{T}_1+1 }
- \frac{ \mathcal{T}_3 +1   }{  \mathcal{T}_2 +1 }
- \frac{\mathcal{T}_5+1 }{ \mathcal{T}_4+1 }
\right\} \frac{\partial^6 \rho}{\partial X^6}.
\end{eqnarray}
In the derivation process of Eqs.~(\ref{KS_trunc11}) and  (\ref{KS_trunc2}), advection terms, namely those terms including  $\rho^2$, are not taken into account for the sake of simplicity.

In order to remove the second term in Eq.~(\ref{KS_trunc11}) for the D1Q7 lattice, the following $\delta f^{eq}_i$ is added to $f^{eq}_i$ of Eq.~(\ref{feq}),
\begin{eqnarray}
\delta f^{eq}_i = \frac{120 \left( \mathcal{T}_1+1 \right) \beta  }{ \left( \mathcal{T}_4+1 \right) \alpha^4} \rho w^{(6)}_i,
\end{eqnarray}
where $w^{(6)}_i$ is defined in Eq.~(\ref{weight_D1Q7}). 

In similar fashion, to remove the fourth derivative term in Eq.~(\ref{KS_trunc2}), the following $\delta \mathcal{M}$ is added to $\mathcal{M}$ in Table~\ref{tab:moment_nonlineareq},
\begin{eqnarray}
\label{delta_M}
\delta \mathcal{M} = 
-\frac{24 \beta^2 \left( \mathcal{T}_1+1 \right) }{\alpha^4  \left( \mathcal{T}_4+1 \right)} 
\left( \frac{\mathcal{T}_2+1 }{2 \left( \mathcal{T}_1+1 \right)} 
- \frac{ \mathcal{T}_3+1 }{\mathcal{T}_4+1  }
\right).
\end{eqnarray}

The remaining error terms in Eqs.~(\ref{KS_trunc11}) and (\ref{KS_trunc2}) are then
\begin{equation}
\label{remained_truc_error_KS}
\small
\left\{\frac{\alpha^4 \left( \mathcal{T}_6+1 \right) }{90  \left( \mathcal{T}_2+1 \right) } 
+\beta \left( \frac{\mathcal{T}_2+1}{\mathcal{T}_1+1 }
-\frac{\mathcal{T}_3 +1}{\mathcal{T}_2 +1}
-\frac{\mathcal{T}_5+1 }{\mathcal{T}_4+1 }\right)
\right\} 
\frac{\partial^6 \rho}{\partial X^6}.  
\small
\end{equation}
If this coefficient of the sixth derivative is positive, the system is very likely to be stable since the coefficient of the fourth derivative is negative.  For $\tau=1$, this condition can be written as
\begin{eqnarray}
\frac{\alpha^4}{90} \ge  \beta.
\label{cond_stability}
\end{eqnarray}
Thus, when  $\beta$ is not sufficiently small, an instability occurs.  When $\tau$ is increased, however, this condition on $\beta$ is weakened, since the coefficient of $\alpha^4$ in Eq.~(\ref{remained_truc_error_KS}) goes as the fourth power of $\tau$, whereas that of $\beta$ goes as the first power of $\tau$ order.  For this reason, increased $\tau$ can enhance the stability. Nevertheless, because the coefficient of the sixth derivative in Eq.~(\ref{remained_truc_error_KS}) monotonically increases with $\tau$, in most cases it is desirable to choose the minimum stable value of $\tau$.

\subsection{Summary of formalism}

For solving the KS equation, the LB equation, Eq.~(\ref{LB_equation_ordinary}), is solved with the following form of $f^{eq}_{i}$ for each lattice speed scheme.

For the D1Q5 lattice, the following form for $f^{eq}_{i}$  is employed,
\begin{equation}
f^{eq}_{i}=\rho \left( w^{(0)}_{i} +  \mathcal{K} w^{(2)}_{i} + \mathcal{M} w^{(4)}_{i}  \right)+ \rho^2 \mathcal{J} w^{(1)}_{i},
\end{equation}
where the weights $w_{i}$ are shown in Eq.~(\ref{weight_D1Q5}), and $ \mathcal{J}$, $\mathcal{K}$,  and $\mathcal{M}$ are shown in Table.~\ref{tab:moment_nonlineareq}. In a previous study~\cite{2017_Otomo}, this model is applied to various test cases and compared with analytic solutions and with the results of LB models used in another study~\cite{2009_Huilin}, whose accuracy was improved~\cite{2017_Otomo}.

For the D1Q7 lattice, the following $f^{eq}_{i}$  is employed,
\begin{eqnarray}
f^{eq}_{i}=\rho \{ w^{(0)}_{i} +  \mathcal{K} w^{(2)}_{i}  + \left( \mathcal{M} + \delta \mathcal{M} \right) w^{(4)}_{i}  \nonumber \\
+\frac{120 \left( \mathcal{T}_1+1 \right) \beta  }{ \left( \mathcal{T}_4+1 \right) \alpha^4}  w^{(6)}_i  \}  + \rho^2 \mathcal{J}  w^{(1)}_{i},
\end{eqnarray}
where the weights $w_{i}$ are shown in Eqs.~(\ref{weight_D1Q5}) and (\ref{weight_D1Q7}), and where $ \mathcal{J}$, $\mathcal{K}$,  and $\mathcal{M}$ are shown in Table.~\ref{tab:moment_nonlineareq}.  Here, $\delta \mathcal{M}$ is defined in Eq.~(\ref{delta_M}). 

%
\section{Comparisons with analytic solutions}
\label{sec:result}

The proposed LB models in Section.~\ref{sec:LB_formalism} are validated by comparing their results with analytic solutions of the KS equation. For quantitative evaluations the global relative error $G$ is defined as,
\begin{equation}
G=\frac{\sum_{x} \left|\rho^{N} \left( x \right) -\rho^{A}\left( x \right)\right|}
{\sum_{x} \left| \rho^{A}  \left( x \right) \right| },
\end{equation}
where $\rho^{N}$ denotes the numerical results, and $\rho^{A}$ the corresponding analytic quantity.


\begin{table*}[htbp]
  \begin{center}
    \caption{ An analytic solution for the Kuramoto-Sivashinsky equation and its initial and boundary conditions, where $X_{M}$ and $X_{m}$ denote the maximum and minimum $X$ coordinates of the domain and $k=\frac{1}{2} \sqrt{\frac{11}{19}}$. In this article, $b=3$, $X_0=\left( X_M-X_m \right)/3$, $X_M=60$, and $X_m=0$.}
    \label{table:KS_analytic_sol}

\begin{tabular}{c c} \hline
      Boundary condition    &  \parbox{12cm}{ \begin{equation} \rho \left(X_{m},t \right)=b-\frac{30}{19} \sqrt{\frac{11}{19}} \end{equation} }   \\
                            &  \parbox{12cm}{ \begin{equation} \rho \left(X_{M},t \right) = 2b-\rho \left(X_{m},t \right) \end{equation} } \\ \hline
     Initial condition     &  \parbox{12cm}{ \begin{equation} \rho \left(X,0 \right) = b+ \frac{15}{19} \sqrt{\frac{11}{19}} \left\{-9 \text{ tanh} \left[k \left(X-X_0 \right)\right]+11 \text{ tanh}^3 \left[k \left(X-X_0 \right) \right]  \right\}  \end{equation} }      \\ \hline
     Analytic solution & \parbox{12cm}{ \begin{equation} \rho \left(X,t \right)=b+\frac{15}{19} \sqrt{\frac{11}{19}} \left\{-9 \text{ tanh} \left[k \left(X-bt-X_0 \right)\right]+11 \text{ tanh}^3 \left[k \left(X-bt-X_0 \right) \right] \right\} \end{equation} } \\ \hline
\end{tabular}
\end{center}
\end{table*}

In Table~\ref{table:KS_analytic_sol}, the analytic solution and its initial and boundary conditions are listed.  Using LB models based on D1Q5 and D1Q7 for the KS equation in Section.~\ref{sec:LB_formalism}, this system was simulated with various $\Delta T$ and $\Delta X$.  As discussed in Section~\ref{sec:KS_formalism}, increasing the relaxation time $\tau$ can improve stability but may cause a decline of precision.  So, for each $\Delta T$, the simulation was run with a variety of $\tau$, and the minimum value required for stability was recorded.  

\begin{figure*}[htbp]
  \begin{center}
    \begin{tabular}{c}
      \begin{minipage}{0.5\hsize}
          \includegraphics[clip, width=7.5 cm]{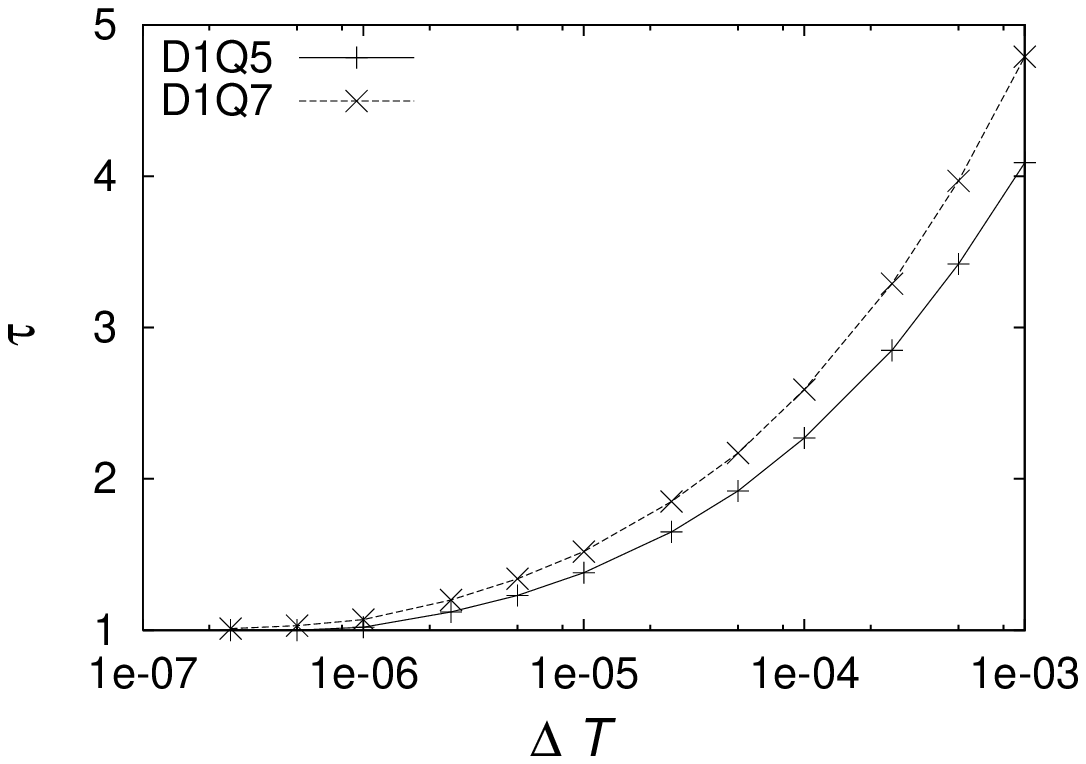}
      \end{minipage}
      \begin{minipage}{0.5\hsize}
        \begin{center}
          \includegraphics[clip, width=7.5 cm]{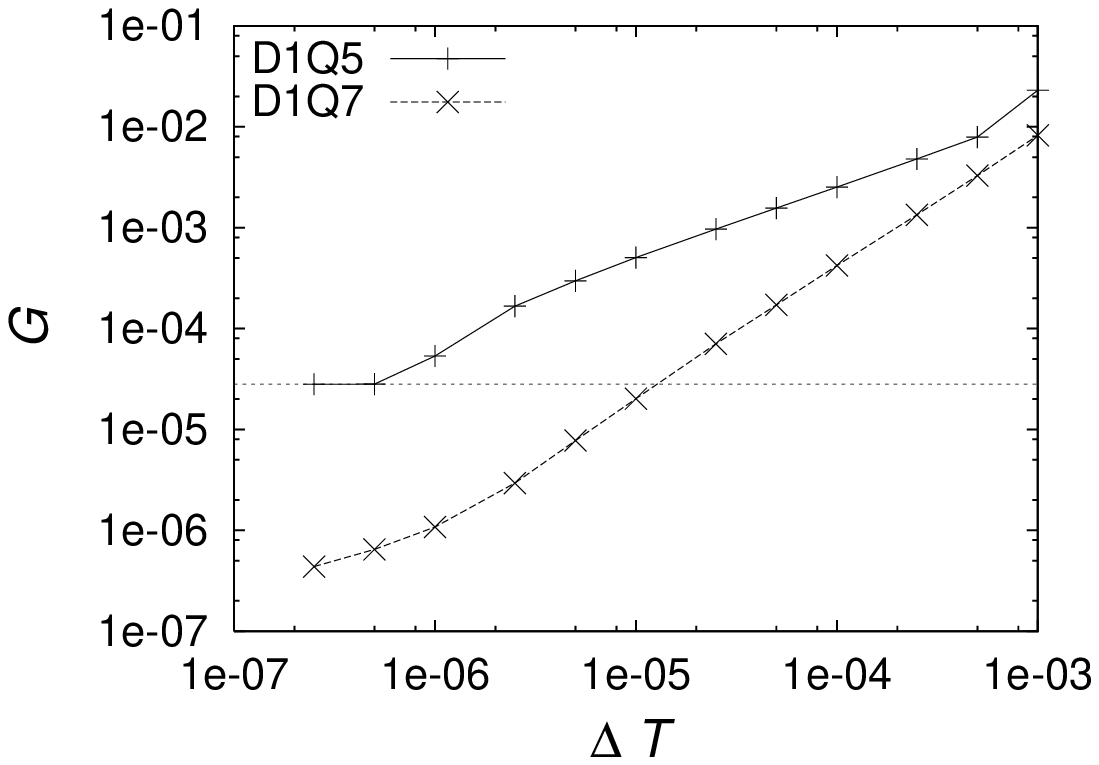}
        \end{center}
      \end{minipage}
    \end{tabular}
    \caption{Minimum relaxation time $\tau$ required for stability (left) and global relative error $G$ (right) as a function of $\Delta T$ for $\Delta X=0.05$ with the basic scheme on the D1Q5 lattice, and the proposed scheme on the D1Q7 lattice. In the right-hand figure, a dotted line is plotted on the plateau region of the D1Q5 model for purposes of comparison with the D1Q7 model.}
    \label{fig:KS_Tau_G_vs_dt_delx005}
  \end{center}
\end{figure*}

\begin{figure*}[htbp]
  \begin{center}
    \begin{tabular}{c}
      \begin{minipage}{0.5\hsize}
          \includegraphics[clip, width=7.5 cm]{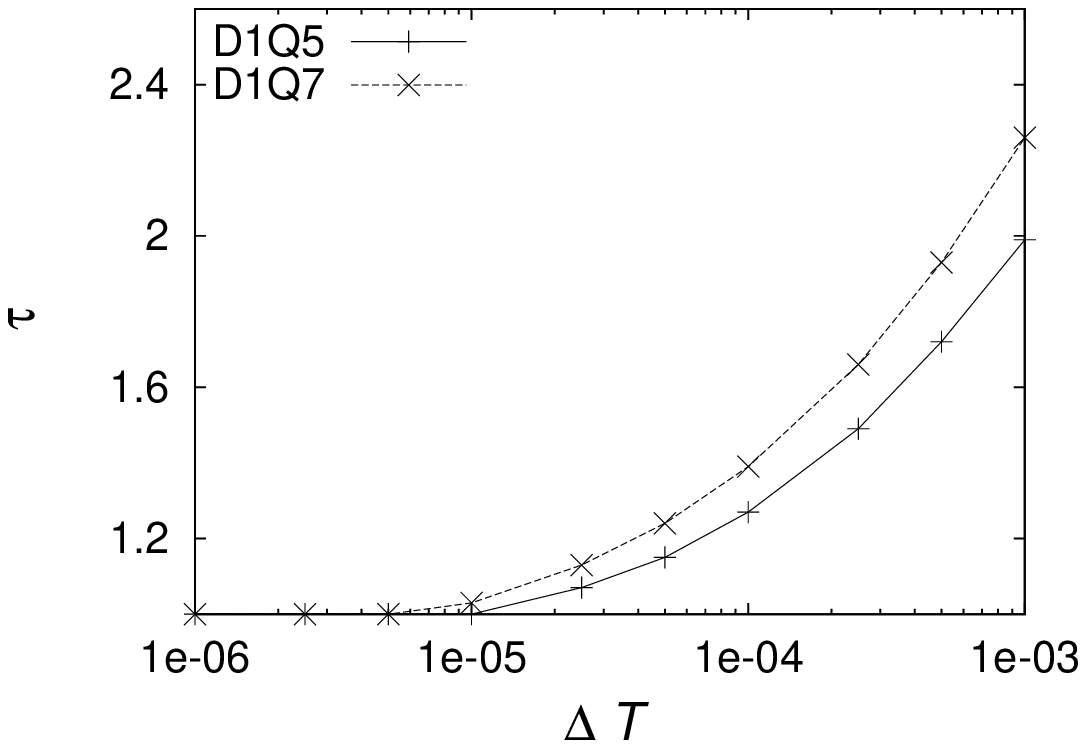}
      \end{minipage}
      \begin{minipage}{0.5\hsize}
        \begin{center}
          \includegraphics[clip, width=7.5 cm]{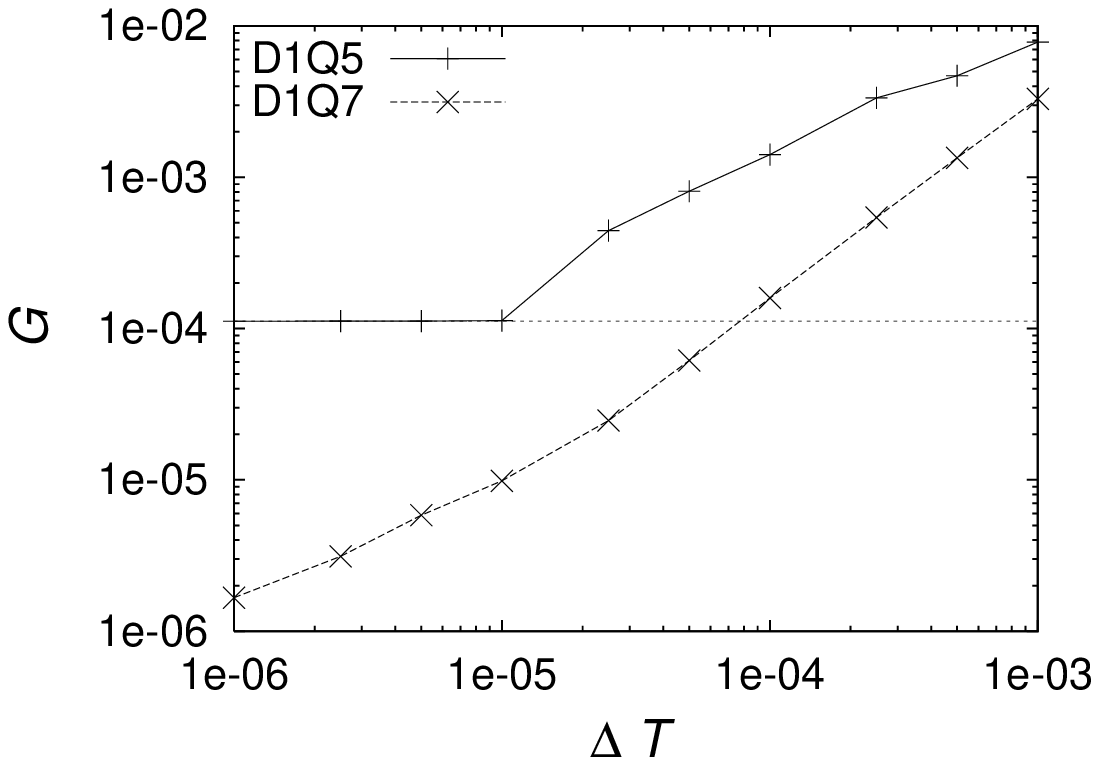}
        \end{center}
      \end{minipage}
    \end{tabular}
    \caption{ Minimum relaxation time $\tau$ required for stability (left) and global relative error $G$ (right) as a function of $\Delta T$ for $\Delta X=0.1$ with the basic scheme on the D1Q5 lattice, and the proposed scheme on the D1Q7 lattice.  In the right-hand figure, a dotted line is plotted on the plateau region of the D1Q5 model for purposes of comparison with the D1Q7 model.}
    \label{fig:KS_Tau_G_vs_dt_delx01}
  \end{center}
\end{figure*}

\begin{figure*}[htbp]
  \begin{center}
    \begin{tabular}{c}
      \begin{minipage}{0.5\hsize}
          \includegraphics[clip, width=7.5 cm]{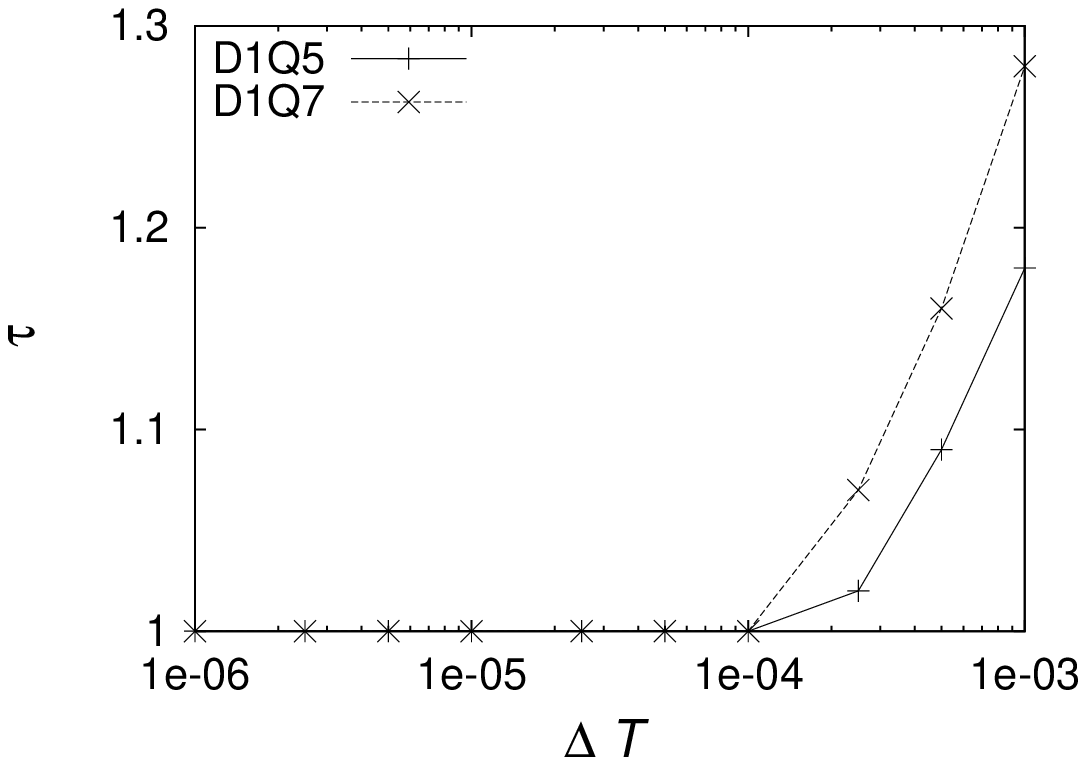}
      \end{minipage}
      \begin{minipage}{0.5\hsize}
        \begin{center}
          \includegraphics[clip, width=7.5 cm]{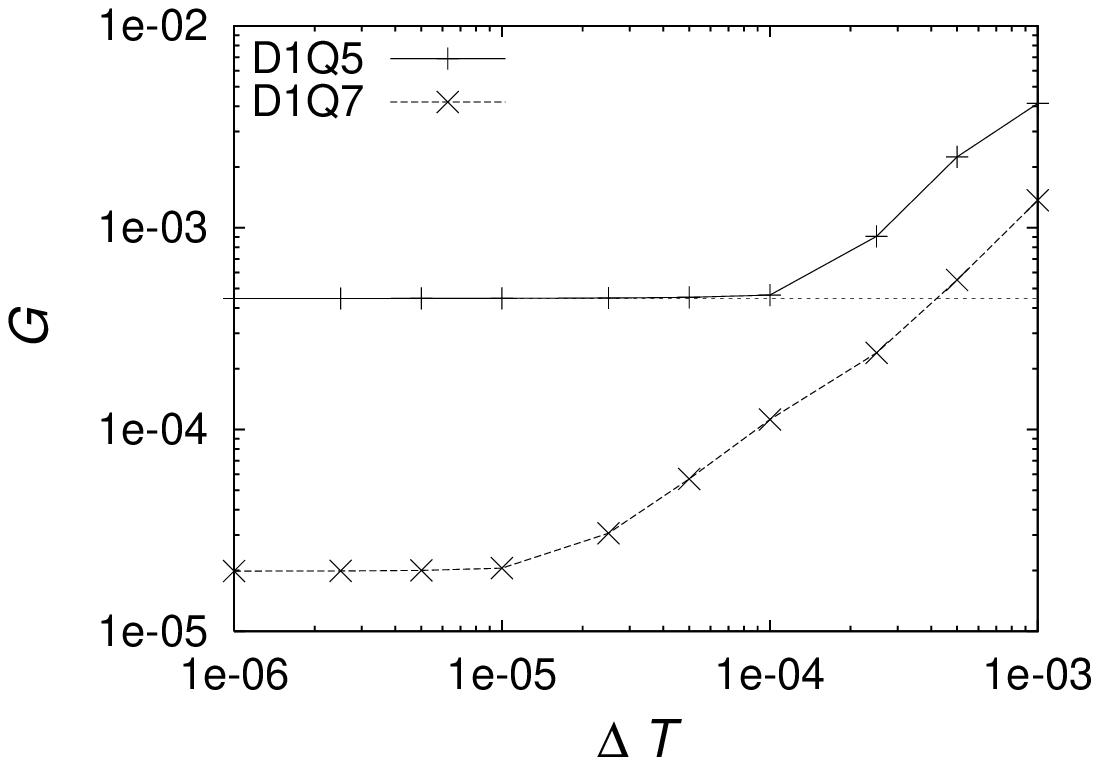}
        \end{center}
      \end{minipage}
    \end{tabular}
    \caption{ Minimum relaxation time $\tau$ required for stability (left) and global relative error $G$ (right) as a function of $\Delta T$ for $\Delta X$=0.2 with the basic scheme on the D1Q5 lattice, and the proposed scheme on the D1Q7 lattice.  In the right-hand figure, a dotted line is plotted on the plateau region of the D1Q5 model for purposes of comparison with the D1Q7 model.}
    \label{fig:KS_Tau_G_vs_dt_delx02}
  \end{center}
\end{figure*}

Figs.~\ref{fig:KS_Tau_G_vs_dt_delx005}, \ref{fig:KS_Tau_G_vs_dt_delx01}, and \ref{fig:KS_Tau_G_vs_dt_delx02} were plotted for $\Delta X =0.05$,  $\Delta X =0.1$, and $\Delta X =0.2$, respectively.  The left-hand plot in each of these figures shows the minimum stable value of $\tau$ versus $\Delta T$ for both the D1Q5 and D1Q7 lattices.  The right-hand plot in each of these figures shows the global relative error $G$ versus $\Delta T$, again for both the D1Q5 and D1Q7 lattices.

According to the discussion in Section~\ref{sec:LB_formalism}, it is clear that, for the D1Q5 LB model, the fourth derivative term in Eq.~(\ref{KS_trunc2}) leads to strong diffusion with increased $\tau$.  By contrast, this term is removed in the D1Q7 LB model.  This demonstrates why the D1Q7 model requires a higher value of $\tau$ for stability than the D1Q5 model, as is evident from Figs.~\ref{fig:KS_Tau_G_vs_dt_delx005} through \ref{fig:KS_Tau_G_vs_dt_delx02}.  

From Figs.~\ref{fig:KS_Tau_G_vs_dt_delx005} through \ref{fig:KS_Tau_G_vs_dt_delx02}, the following expression for the minimum stable $\tau$ as a function of $\Delta X$ and $\Delta T$ can be deduced,
\begin{eqnarray}
\footnotesize
\tau = \begin{cases}
    A \left( log_{10} \left( \Delta T \right) +B \right)^2 +1 & \mbox{for $\log_{10} \left( \Delta T \right) \ge -B$} \\
    1 & \mbox{otherwise.}
  \end{cases}
\footnotesize
\label{general_tau_formula}
\end{eqnarray}
Here $A$ and $B$ are functions of $\Delta X$.  By fitting the results in Figs.~\ref{fig:KS_Tau_G_vs_dt_delx005} through \ref{fig:KS_Tau_G_vs_dt_delx02}, one obtains for the D1Q5 LB model
\begin{eqnarray}
\label{A_B_fit_D1Q5}
A &=& 0.103 \Delta X^{-0.381} \nonumber \\
B &=& -13.2 \Delta X + 6.56,
\end{eqnarray}
and for the D1Q7 LB model
\begin{eqnarray}
\label{A_B_fit_D1Q7}
A &=& 0.0886 \Delta X^{-0.472} \nonumber \\
B &=& -12.5 \Delta X + 6.64.
\end{eqnarray}
If the case of $\Delta X$=0.1 is considered, the conditional branches of $\Delta T$ in Eq.~(\ref{general_tau_formula}) are $\Delta T_{crit}=5.75\times 10^{-6}$ for the D1Q5 model, and $\Delta T_{crit}=4.07\times 10^{-6}$ for the D1Q7 model, which are computed by $B$ in Eqs.~(\ref{A_B_fit_D1Q5}) and  Eqs.~(\ref{A_B_fit_D1Q7}).  
$\Delta T_{crit}$ shows maximum time increments in order to keep stability with $\tau=1$ and thereby can be regarded as a criteria of stability.
Comparing with the maximum $\beta$, i.e., $\Delta T$,  in Eq.~(\ref{cond_stability}), whose value is $1.11\times 10^{-6}$, one confirms that Eq.~(\ref{cond_stability}) is a sufficient stability condition. 
Moreover since $\Delta T_{crit}$ in the D1Q7 model is slightly smaller than that for the D1Q5 model, stability when $\tau$=1 is worse for D1Q7, but by increasing $\tau$ this drawback is removed.

According to the results for the D1Q5 model, when $\Delta T$ is below a certain value, $G$ is insensitive to $\Delta T$.  This is probably because the dominant error is coming from terms of order $\beta^0$, namely Eq.~(\ref{KS_trunc11}). 
Indeed, this plateau region is at lower $\Delta T$ if $\Delta X$ is smaller, because $\Delta X$, namely $\alpha$, includes the error shown in Eq.~(\ref{KS_trunc11}).
And according to Fig.~\ref{fig:KS_Tau_G_vs_dt_delx02}, the plateau region of D1Q7 is at lower $\Delta T$ than that of D1Q5, since we removed the second term in Eq.~(\ref{KS_trunc11}) for the D1Q7 model. 

\begin{figure*}[htbp]
  \begin{center}
    \begin{tabular}{c}
      \begin{minipage}{0.5\hsize}
          \includegraphics[clip, width=7.5 cm]{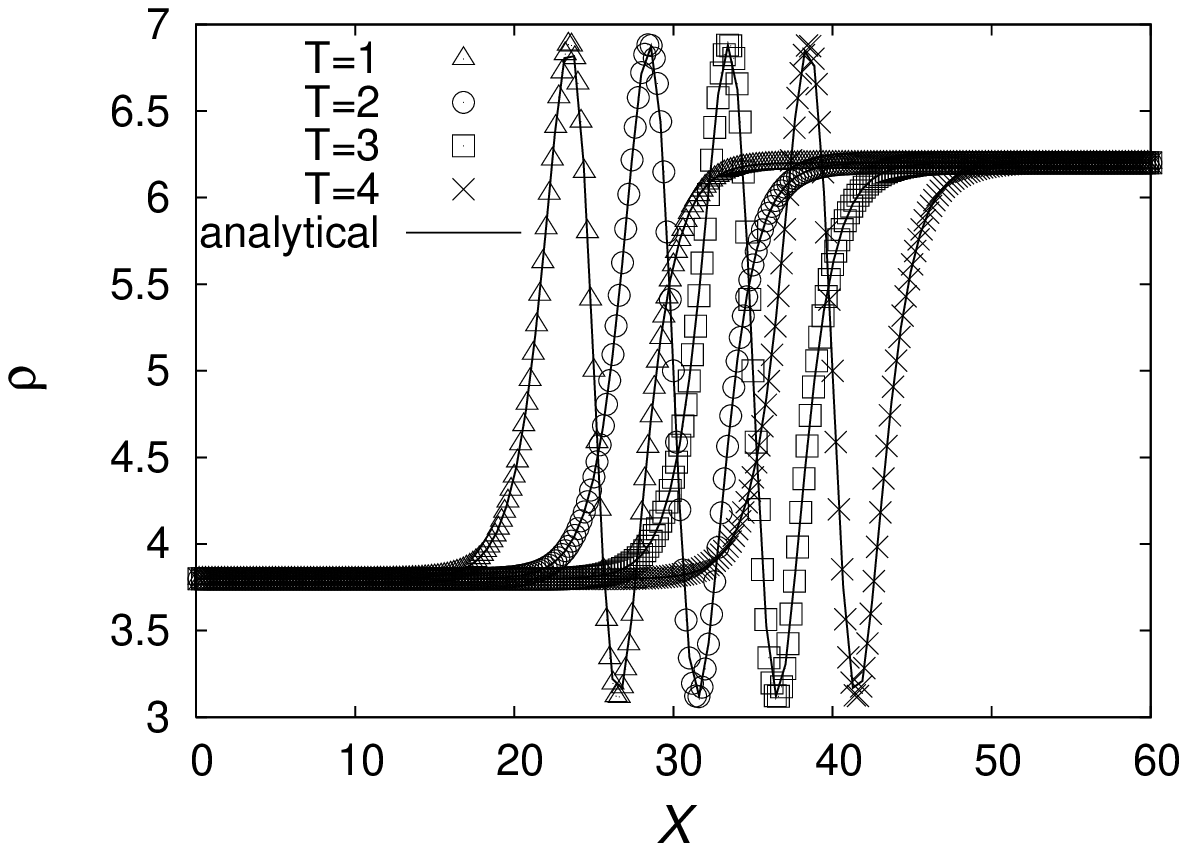}
      \end{minipage}
      \begin{minipage}{0.5\hsize}
        \begin{center}
          \includegraphics[clip, width=7.5 cm]{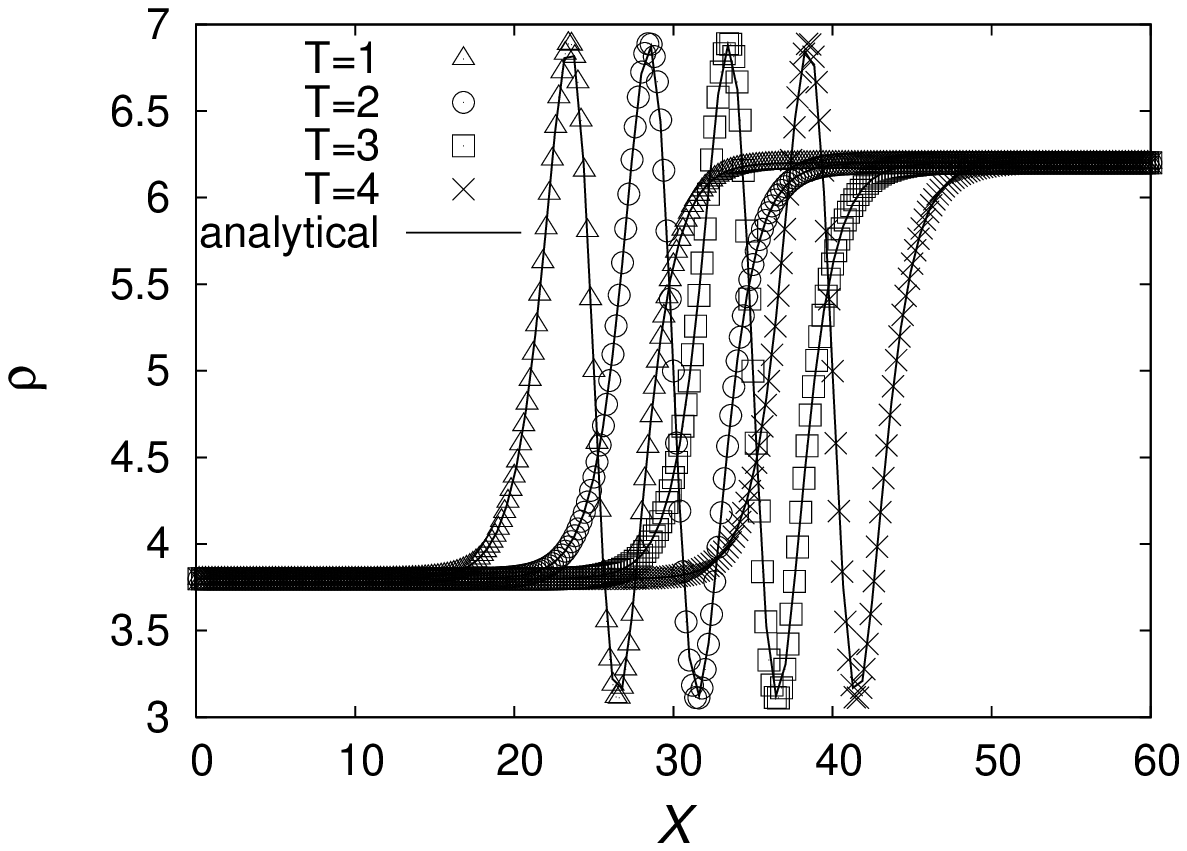}
        \end{center}
      \end{minipage}
    \end{tabular}
    \caption{ Comparisons with the analytic solution for each $T$ using the basic D1Q5 scheme (left) and the proposed D1Q7 scheme (right) for $\Delta X$=0.2. }
    \label{fig:KS_density_comparison}
  \end{center}
\end{figure*}

In order to compare accuracy between the D1Q5 and D1Q7 models, dotted lines are drawn on the plateau regions for the D1Q5 model in the right-hand graphs of Figs.~\ref{fig:KS_Tau_G_vs_dt_delx005} through \ref{fig:KS_Tau_G_vs_dt_delx02}.  The crossed points show corresponding results between the D1Q5 and D1Q7 models with the same accuracy level. Computational costs, measured by averaging over five trials with an Intel Xeon 3.5GHz core, and profiles of $\rho$ versus $X$ for $\Delta X = 0.2$ are shown in Table~\ref{table:comput_cost_KS} and Fig.~\ref{fig:KS_density_comparison}. In Fig.~\ref{fig:KS_density_comparison}, the numerical results agree very well with the analytic solution for any $T$, and for both the D1Q5 and D1Q7 models, even at coarse resolution such as $\Delta X=0.2$.
Table.~\ref{table:comput_cost_KS} shows that the proposed scheme on the D1Q7 lattice saves computational cost by $65 - 92  \%$ compared to the basic scheme on the D1Q5 lattice, and this improvement is even more significant as spatial resolution is increased.

\begin{table*}
  \begin{center}
\begin{tabular}{c c c} \hline 
        $\Delta X$ &  D1Q5  &  D1Q7  \\ \hline
          0.05     &     4604.25  $\pm$  4.26 &  357.75 $\pm$ 0.43          \\
          0.1      &     110.75 $\pm$ 0.43 & 18.00 $\pm$ 0.00           \\
          0.2      &     5.75 $\pm$ 0.43 & 2.00 $\pm$ 0.00           \\ \hline
\end{tabular}
\caption{ Comparisons of computational costs (sec) with the basic scheme on D1Q5 and the proposed scheme on D1Q7  in terms of various $\Delta X$ using corresponding $\Delta T$ which yields the same accuracy level}
\label{table:comput_cost_KS}
\end{center}
\end{table*}

%
\section{SUMMARY}
\label{sec:summary}
%

Using the Taylor-series expansion method, higher-order effects of LB models for the Kuramoto-Sivashinsky equation have been analyzed, both for the D1Q5 and D1Q7 lattices.
It was found that the proper choice of the relaxation time $\tau$ enhances stability, and that the optimized scheme with the D1Q7 lattice improves the accuracy.
These two features are complementary.  In particular, enhancements of stability by increasing $\tau$, which allows simulation with larger time increments, compensates for the additional computational costs associated with a larger number of lattice speeds.
Moreover, the improvement of accuracy by including more lattice speeds compensates for the deteriorated precision due to increased $\tau$. 
As a consequence of all these effects, the computational costs of the proposed D1Q7 LB scheme is reduced by up to $94\%$ compared to the original LB model on the D1Q5 lattice.
In the future, using this scheme, we plan to use these LB models to elucidate the dynamics of this very interesting chaotic dynamical system.

\section*{Acknowledgements}

This work was supported by a public grant from the Fondation Math\'ematique  Jacques Hadamard as part of the "Investissement d'avenir" project, reference ANR-11-LABX-0056-LMH, LabEx LMH.

%
\appendix
\section{Weights in the equilibrium state}
\label{appendix:weight}
%
In this Appendix, for both the D1Q5 and D1Q7 models, a discrete set of weights having only the unit $l$th moment, $w^{(l)}_i$, is presented.  This set of weights has the property
\begin{eqnarray}
\label{cond_weight}
\sum_{i} c^{p}_i w^{(l)}_i = \delta_{p,l},
\end{eqnarray}
where $\delta$ denotes the Kronecker delta, and where $p$ ranges from 0 to one less than the number of velocities.  

For the D1Q5 model, with $c_i= \left\{ 0, \pm 1, \pm 2 \right\}$, the required set of $w^{(l)}_i$ can be obtained by inverting the matrix;
\begin{eqnarray}
\begin{pmatrix} 
1 & 1 & 1 & 1 & 1 \\
0 & 1 & -1 & 2 & -2 \\
0 & 1 &  1 & 4 &  4 \\
0 & 1 &  -1 & 8 &  -8 \\
0 & 1 &  1 & 16 &  16 
\end{pmatrix}.
\end{eqnarray}
The result is
\begin{eqnarray}
\label{weight_D1Q5}
\begin{pmatrix} w^{(0)}_i \\ w^{(1)}_i \\ w^{(2)}_i \\ w^{(3)}_i \\ w^{(4)}_i \end{pmatrix}
=
\begin{pmatrix}
\left\{ 1, 0 ,0\right\} \\ \left\{ 0, \pm \frac{2}{3} , \mp \frac{1}{12} \right\} \\
\left\{ -\frac{5}{4},  \frac{2}{3} , - \frac{1}{24} \right\} \\ \left\{ 0, \mp \frac{1}{6} , \pm \frac{1}{12} \right\} \\  \left\{ \frac{1}{4}, - \frac{1}{6} ,  \frac{1}{24} \right\}
\end{pmatrix}.
\end{eqnarray}

For the D1Q7 model with $c_i= \left\{ 0, \pm 1, \pm 2 , \pm 3 \right\}$, the required set of $w^{(l)}_i$ can be obtained by inverting the matrix;
\begin{eqnarray}
\begin{pmatrix} 
1 & 1 & 1 & 1 & 1 & 1  & 1 \\
0 & 1 & -1 & 2 & -2 & 3 & -3 \\
0 & 1 &  1 & 4 &  4 & 9 &  9 \\
0 & 1 &  -1 & 8 & -8 & 27  & -27 \\
0 & 1 &  1 & 16 &  16 & 81 & 81 \\
0 & 1 &  -1 & 32 & -32 & 243 & -243 \\
0 & 1 &   1 & 64 &  64 & 729 &  729
\end{pmatrix}.
\end{eqnarray}
The results for $l=5$ and $l=6$ are
\begin{eqnarray}
\label{weight_D1Q7}
\begin{pmatrix} w^{(5)}_i \\ w^{(6)}_i  \end{pmatrix}
=
\begin{pmatrix}
\left\{ 0, \pm \frac{1}{48} , \mp \frac{1}{60} , \pm \frac{1}{240} \right\}  \\
\left\{ -\frac{1}{36},  \frac{1}{48} ,  -\frac{1}{120}, \frac{1}{720} \right\}
\end{pmatrix}.
\end{eqnarray}

\end{document}